\documentclass{article}

\usepackage{arxiv}

\usepackage[utf8]{inputenc} 
\usepackage[T1]{fontenc}    
\usepackage{hyperref}       
\usepackage{url}            
\usepackage{booktabs}       
\usepackage{amsfonts}       
\usepackage{nicefrac}       
\usepackage{microtype}      
\usepackage{lipsum}		
\usepackage{graphicx}
\usepackage{natbib}
\usepackage{doi}

\usepackage{amsmath}
\usepackage{adjustbox}
\usepackage{amssymb}
\usepackage{amsthm}
\usepackage{psfrag,epsf}
\usepackage{enumerate}
\usepackage{graphics}
\usepackage[english]{babel}
\usepackage[utf8]{inputenc}
\usepackage[colorinlistoftodos]{todonotes}
\usepackage{float}
\usepackage{bm}
\usepackage{diagbox}
\usepackage{mathtools}
\usepackage{multirow}
\usepackage{titlesec}
\usepackage{xr}
\usepackage{lipsum}
\usepackage{caption}
\usepackage{subcaption}
\usepackage{chngcntr}

\newcommand{\argmin}[1]{\underset{#1}{\operatorname{arg}\,\operatorname{min}}\;}

\newcommand*{\X}{\mathbf{X}}

\newcommand*{\x}{\mathbf{x}}
\newcommand*{\y}{\mathrm{y}}
\newcommand*{\N}{\mathcal{N}}
\newcommand*{\m}{\mathbf{m}}

\providecommand{\keywords}[1]
{
	\small
	\textbf{\textit{Keywords---}} #1
}

\title{Dirichlet Process Mixture Models\\ with Shrinkage Prior}

\date{} 					

\author{ Dawei Ding \\
	Department of Mathematics, Statistics, and Computer Science \\
	University of Illinois at Chicago\\
	Chicago, IL 60015 \\
	\texttt{dding20@uic.edu} \\
	\And
	George Karabatsos \\
	Department of Educational Psychology, \\ 
	with affiliation to the Department of Mathematics, Statistics, and Computer Science \\
	University of Illinois at Chicago\\
	Chicago, IL 60015 \\
	\texttt{gkarabatsos1@gmail.com} \\
}

\renewcommand{\shorttitle}{DPM Models with Shrinkage Prior}

\hypersetup{
pdftitle={Dirichlet Process Mixture Models with Shrinkage Prior},
pdfsubject={},
pdfauthor={Dawei Ding, George Karabatsos},
pdfkeywords={Bayesian nonparametrics, Shrinkage Prior, Regression, Variable selection},
}

\begin{document}
\maketitle

\begin{abstract}
	We propose Dirichlet Process Mixture (DPM) models for prediction and cluster-wise variable selection, based on two choices of shrinkage baseline prior distributions  for the linear regression coefficients, namely the Horseshoe prior and Normal-Gamma prior. We show in a simulation study that each of the two proposed DPM models tend to outperform the standard DPM model based on the non-shrinkage normal prior, in terms of predictive, variable selection, and clustering accuracy. This is especially true for the Horseshoe model, and when the number of covariates exceeds the within-cluster sample size. A real data set is analyzed to illustrate the proposed modeling methodology, where both proposed DPM models again attained better predictive accuracy.
\end{abstract}


\section{Introduction}
\label{intro}
For linear regression with variable (covariate) selection, the LASSO provides a prominent method 
with many extensions \citep{tibshirani1996regression, tibshirani2011regression}. This method employs a shrinkage parameter which can shrink the regression coefficients of irrelevant covariates to zero, and the corresponding LASSO estimate can be interpreted as the Bayes posterior mode under independent zero-mean Laplace prior distributions on the regression coefficients. \cite{park2008bayesian} first considered a fully Bayesian LASSO by exploiting the
representation of the Laplace prior as a scale mixture of zero-mean normal
distributions with exponential mixing density \citep{andrews1974scale}. However, such a choice of prior does not provide adaptive shrinkage, but instead shrinks all coefficients towards zero. Other shrinkage priors, defined by other mixing distributions, were proposed mainly to address this issue. \cite{carvalho2009handling, carvalho2010horseshoe} proposed the Horseshoe prior, defined by a half-Cauchy mixing distribution, with local shrinkage parameters which help achieve robustness in handling sparsity. \cite{griffin2010inference} proposed the Normal-Gamma shrinkage prior, defined by a Gamma mixing distribution which provides adaptive tail thickness and shrinkage. These previous studies showed that, for normal linear models, models assigned the Horseshoe or Normal-Gamma prior on the coefficients tend to outperform models assigned a Laplace (LASSO) or non-shrinkage normal prior on the coefficients, in terms of parameter estimation and variable selection accuracy, especially when the number of covariates exceeds the sample size. Furthermore, statistical estimation with models based on continuous shrinkage priors, such as the Horseshoe, Normal-Gamma, and Laplace (LASSO), can be more computationally efficient than estimation with models based on spike-and-slab priors, because the latter models can require high computational cost to search through a large number of possible
submodels when there are many covariates \citep{castillo2015bayesian}.

One key limitation of the existing shrinkage priors is that they do not allow variable selection to vary over different clusters of the data points, based on the unknown clustering estimated from the data. However, few methods have addressed this problem; see \cite{barcella2017comparative} for a review. \cite{barcella2016variable} proposed a covariate-dependent Dirichlet Process Mixture (DPM) model which assigns a spike-and-slab prior distribution to achieve cluster-wise variable selection for binary covariates. \cite{quintana2015cluster} proposed cluster-wise variable selection in a
product partition model by employing binary indicator parameters on the covariate similarity function. Meanwhile, the joint DPM modeling approach, defined by DPM modeling of both covariate and conditional response distributions, can provide better predictive accuracy than DPM modeling which assumes fixed covariates. This is because the approach accounts for the distance between a new covariate profile $\x$ and the data-observed $\x$ values of the different cluster groups \citep{hannah2011dirichlet}. 

To address the key limitation, we propose two joint DPM of regression models, defined by either a Horseshoe or Normal-Gamma shrinkage baseline prior on the regression coefficients. Each shrinkage DPM model is based on the Dependent Dirichlet Process (DDP), a wide and flexible
class of covariate-dependent random probability measures \citep{maceachern1999dependent, quintana2020dependent}. Next, in Section 2, we further describe our proposed DPM models. It also characterizes the parameter posterior distributions and conditional posterior predictive distributions, and provides corresponding MCMC sampling algorithms for the models. This section also reviews methods for summarizing the MCMC posterior sample output 
for variable selection and clustering estimation. In Section 3, we compare our DPM models against the standard DPM model which 
assigns non-shrinkage normal baseline prior on the regression coefficients, in terms of predictive, variable selection, and clustering accuracy. 
Section 4 illustrates our shrinkage DPM models on a real data set. Section 5 concludes the article.

\section{Model and Posterior Inference}

Given a data set matrix $(\bm{\y},\X)=(y_i, \x_i)_{i=1}^n$ including $n$ observations of $p \times 1$
covariate vectors $\x_i \in \mathbb{R}^p$ and scalar responses $y_i \in \mathbb{R}$, our
DPM model is the mixture of Normal ($\N$) probability densities: 
\begin{align}  \label{model_1}
(y_i,\x_i) | G, \sigma^2 \overset{iid}{\sim} \int \N(y|\mu+\x^T\bm{\beta},\sigma^2)
\N_p(\x| \m,\text{diag}(\bm{\tau})) G(d\bm{\theta}),
\end{align}
for $i=1,\dots,n$, where $\bm{\theta} = \{\bm{\theta} _{y}, \bm{\theta} _{\x%
}\}$, $\bm{\theta} _{y} = \{\mu,\bm{\beta}\}$, $\bm{\theta} _{\x} = \{\m%
,\bm{\tau}\}$, $\bm{\beta} = (\beta_1,\dots,\beta_p)$, $\m = (m_1,\dots,m_p)$, and $\bm{\tau} = (\tau_1,\dots,\tau_p)$ with mixing distribution $G \sim \text{DP}(\alpha, G_0)$ assigned a Dirichlet Process prior with mass parameter $\alpha>0$ and
baseline measure $G_0$ \citep{ferguson1973bayesian}. The precision parameter $\alpha$ is assigned a gamma $\mathit{Ga}(c,d)$ prior distribution with shape parameter $c$ and rate parameter $d$.

Since the support of the DP prior is almost surely discrete, the general
mixture model (\ref{model_1}) can also be expressed as the countable mixture:
\begin{align}  \label{model_stickbreak}
f(y,\x|G ,\sigma^2) = \sum\limits_{j=1}^{\infty} \N(y|\mu_j+\x^T\bm{\beta}%
_j,\sigma^2) \N_p(\x|\m_j,\text{diag}(\bm{\tau}_j)) w_j.
\end{align}
The Dirichlet Process $G$ in (\ref{model_1}) admits the stick-breaking representation
of infinite mixture $G(\cdot)=\sum\limits_{j=1}^{\infty} w_j \delta_{\bm{\theta}_j}(\cdot)$, where $\bm{\theta}_j = \{\mu_j, \bm{\beta}_j, \m_j, \bm{\tau}_j\} \overset{iid}{\sim} G_0$,  $\delta_{\bm{\theta}_j}$ is the Dirac measure
that takes the value 1 on $\bm{\theta}_j$ and 0 elsewhere, and the mixing weights have the stick-breaking
form $w_j=v_j \prod\limits_{l=1}^{j-1}(1-v_l)$ with $v_j \stackrel{iid}{\sim} \mathit{Beta}(1,\alpha)$ for $j=1,2,\dots$, $\sum\limits_{j=1}^{\infty} w_j
= 1$ \citep{sethuraman1994constructive}.

The conditional density of $Y$ in the joint mixture model (\ref{model_stickbreak}) can be written as:
\begin{align*}
f(y|\x, G, \sigma^2) = \sum\limits_{j=1}^{\infty} \N(y|\mu_j+\x^T\bm{\beta}%
_j,\sigma^2) w_j(\x), 
\end{align*}
with covariate dependent mixture weights $w_j(\x) = \frac{w_j \N_p(\x|\m_j,\text{diag}(\bm{\tau}_j))}{\sum\limits_{h=1}^{\infty} w_h \N_p(\x|\m_h,\text{diag}(\bm{\tau}_h))}$, implying that the model is based on the Dependent Dirichlet Process \citep{quintana2020dependent}.

\subsection{Horseshoe DPM Model (HS-DPM)}
The HS-DPM model is completed by the specification of the following prior distributions, while defining the baseline measure $G_0$ according to the Horseshoe prior \citep{carvalho2010horseshoe} for the regression coefficients in the infinite component representation (\ref{model_stickbreak}):
\begin{subequations}
	\label{hs_prior}
	\begin{align}
	& g_0(\mu_j, \bm{\beta}_j, \m_j, \bm{\tau}_j) = \N(\mu_j|0,\nu_{\mu}) \N_p(
	\bm{\beta}_j | \bm{0}, \zeta^2_j \sigma^2 \bm{\Gamma}_j) 
	\prod\limits_{l=1}^p \Big[ \N(m_{jl}|m_0,\frac{\tau_{jl}}{n_0}) \mathit{IG}(\tau_{jl}|\frac{\nu_0}{2}, \frac{2}{\nu_0 s_0^2}) \Big]  \\
	& \bm{\Gamma}_j = \text{diag}(\gamma_{j1}^2,\dots,\gamma_{jp}^2) \equiv \text{
		diag}(\bm{\gamma}^2_j), \ j=1,2,\dots  \\
	& \pi(\bm{\gamma}_j) = \prod \limits_{l=1}^{p} \mathcal{C}^{+}(\gamma_{jl} | 0,1), \ j=1,2,\dots \\
	& \pi(\zeta_j)  = \mathcal{C}^{+}(\zeta_j | 0,1), \ j=1,2,\dots \\
	& \pi(\sigma^2) = \mathit{IG}(\sigma^2 | \alpha_0, \theta_0).
	\end{align}
\end{subequations}
Above, $g_0$ is the pdf of the baseline measure $G_0$, $\mathit{IG}(c,d)$ is the Inverse-gamma distribution with shape $c$ and scale $d$, and $\mathcal{C}^+(0,1)$ is the standard half-Cauchy distribution.

\subsection{Normal-Gamma DPM Model (NG-DPM)}
\label{ng-dpm}
The NG-DPM model is completed by the specification of the following prior density functions for the model parameters, while defining the baseline measure $G_0$ according to the Normal-Gamma prior \citep{griffin2010inference} for the regression coefficients in the infinite component representation (\ref{model_stickbreak}):

\begin{subequations}
	\label{ng_prior}
	\begin{align}
	& g_0(\mu_j, \bm{\beta}_j, \m_j, \bm{\tau}_j) = \N(\mu_j|0,\nu_{\mu}) \N_p(
	\bm{\beta}_j | \bm{0}, {\bm{D}_{\psi}}_j)  \label{2.3a} 
	\prod\limits_{l=1}^p \Big[ \N(m_{jl}|m_0,\frac{\tau_{jl}}{n_0}) \mathit{IG}(\tau_{jl}|\frac{\nu_0}{2}, \frac{2}{\nu_0 s_0^2}) \Big] \\
	& {\bm{D}_{\psi}}_j = \text{diag}(\psi_{j1},\dots,\psi_{jp}) \equiv \text{
		diag}(\bm{\psi}_j), \ j=1,2,\dots  \\
	& \pi(\bm{\psi}_j | \lambda_j,\gamma^{-2}_j) = \prod\limits_{l=1}^p \mathit{Ga}(\psi_{jl}|\lambda_j,\frac{1}{2}\gamma^{-2}_j), \ j=1,2,\dots \\
	& \pi(\lambda_j, \gamma^{-2}_j) = \mathit{Exp} (\lambda_j | 1) \mathit{Ga}
	(\gamma^{-2}_j|2,\frac{2 V}{\lambda_j}), \ j=1,2,\dots \\
	& \pi(\sigma^2) = \mathit{IG}(\sigma^2 | \alpha_0, \theta_0), 
	\end{align}
\end{subequations}
where $\mathit{Exp}(1)$ is the Exponential distribution with rate 1, $V = \frac{1}{p} \sum\limits_{l=1}^p \hat{\beta}_{l}^2 \bm{1}(n \geq p+1) + \frac{1}{n} \sum\limits_{l=1}^p \tilde{\beta}_{l}^2 \bm{1}(n < p + 1)$, $\bm{1}(\cdot)$ is the indicator function, $(\hat{\mu}, \hat{\bm{\beta}}^T)^T = ({\tilde{\X}}^T \tilde{\X})^{-1} \tilde{\X}^T \bm{\y}$ is the least squares estimate, $(\tilde{\mu}, \tilde{\bm{\beta}}^T)^T = {\tilde{\X}}^T (\tilde{\X} {\tilde{\X}}^T)^{-1} \bm{\y}$ is the minimum
norm least squares estimate, and $\tilde{\X}=[\bm{1}_n, \X]$.

For the HS-DPM and the NG-DPM model, the joint prior density of all the model parameters $\bm{\Lambda} = (\bm{\theta}, \sigma^2)$ in model (\ref{hs_prior}) or (\ref{ng_prior}) is denoted by $\pi(\bm{\Lambda})$, respectively.

\subsection{Posterior Computations}
In order to enable tractable posterior-based inferences, the infinite dimensionality of the HS-DPM model or NG-DPM model can be handled by the introduction of latent variables $u_i$ and cluster membership indicators $d_i$ for $i=1,\dots,n$. Then, for either model, it can be shown that the joint posterior distribution of all the model parameters $\bm{\Lambda}$ and $\{(u_i, d_i),i=1,\dots,n\}$, is proportional to:
\begin{align}
\label{model:joint}
& \prod\limits_{i=1}^n \Big[ \bm{1}(u_i < w_{d_i}) \N(y_i| \mu_{d_i}+\x^T 
\bm{\beta}_{d_i}, \sigma^2) \prod\limits_{l=1}^p \N(x_{il}|m_{d_i,l},\tau_{d_i,l}) \Big] \pi(\bm{\Lambda})
\end{align}
where $d_{i}=j$ if observation pair $(y_{i},\x_{i})$ belongs to $j$th cluster, for $j=1,2,\dots$. Posterior inference of the DPM model proceeds after marginalizing over the latent variables $\{u_i: i=1,\dots,n\}$.

Posterior inference with the HS-DPM or NG-DPM model can be undertaken by using an MCMC sampling algorithm, which embeds the Gibbs sampling algorithm for normal linear models assigned a Horseshoe prior \citep{makalic2015simple}, or assigned a Normal-Gamma prior \citep{griffin2010inference}, within a slice sampler for DPM regression models \citep{karabatsos2012bayesian}. The Appendix provides more details.

\subsection{Posterior Predictive Inference}
Posterior-based prediction from the HS-DPM or NG-DPM model is based on a generalized P{\'o}lya urn scheme \citep{hannah2011dirichlet,wade2014improving}, described as follows. A clustering of the $n$ data points is denoted as $\rho_n=\{d_i\}_{i=1}^n$, comprised of $K$ distinct clusters or values of the $d_i$.
Denote the corresponding cluster sets and members 
as $C_j=\{i: d_i=j\}$ and $\X_j^*=\{\x_i:i\in C_j\}, \bm{\y}_j^*=\{y_i:i\in C_j\}$, for $j=1,\dots,K$. 
Then, based on a covariate-dependent P{\'{o}}lya urn scheme, the cluster label for a
new subject $d_{n+1}$ conditionally on its corresponding profile $\x$, current
clustering $\rho _{n}$, and observations $\X$, has distribution given by:
\begin{align}
\label{cluster_allocation}
d_{n+1}|\x,\X,\rho _{n}\sim \frac{\alpha \pi _{n}}{b_{0}}f_{0,\x}(\x%
)\delta _{K+1}(\cdot )+\frac{1}{b_{0}}\sum\limits_{j=1}^{K}\pi _{n}n_{j}f_{j,%
	\x}(\x)\delta _{j}(\cdot ),
\end{align}%
where $\pi_n = \frac{1}{\alpha + n}$, $n_{j}=|C_j|$, $\sum%
\limits_{j=1}^{K}n_{j}=n$, $b_{0}=p(\x|\rho _{n},\X)$,
\begin{align*}
& f_{0,\x}(\x)=\int \N_p(\x|\bm{\theta}_{\x})G_{0}(d\bm{\theta}_{\x}),\text{ and } f_{j,\x}(\x)=\int \N_p(\x|\bm{\theta}_{\x})p(\bm{\theta}_{\x}|\X_j^*)d\bm{\theta}_{\x}.
\end{align*}%
Thus, the cluster allocation probability distribution (\ref{cluster_allocation}) depends on new $\x$ and the covariate data $\X$. The more similar a new profile $\x$ is to the existing $\x_{i}$'s in
cluster $j$, the higher the density $p(\x|\bm{\theta}_{\x})$, leading to
higher predictive density $f_{j,\x}(\x)$, thus the higher probability of
allocation to cluster $j$.

Once we obtain the allocation scheme for a new subject, we can then derive
the conditional predictive density of the subject's response $y$ for fixed variance $\sigma^2$, 
given her new profile $\x$ and the data $(\bm{\y}, \X)$, as:
\begin{align*}
& f(y | \bm{\y}, \X,  \x) =\sum_{\rho_n} \sum_{d_{n+1}} f(y | \bm{\y}, \X, \x, \rho_n,
d_{n+1}) p(d_{n+1} | \X, \x, \rho_n) p(\rho_n | \bm{\y}, \X, \x) \\
& = \sum_{\rho_n} \Big[ \frac{\alpha \pi_n}{b_0} f_{0,\x}(\x) f_{0,y}(y|\x) + \frac{1}{b_0} \sum\limits_{j=1}^K n_j \pi_n f_{j,\x}(\x) f_{j,y}(y|\x) \Big] \frac{p(\x | \rho_n, \X) p(\rho_n | \bm{\y}, \X)}{p(\x | \bm{\y}, \X)} \\
& = \sum_{\rho_n} \Big[ \frac{\alpha \pi_n}{b_0} f_{0,\x}(\x) f_{0,y}(y|\x) + \frac{1}{b_0} \sum\limits_{j=1}^K n_j \pi_n f_{j,\x}(\x) f_{j,y}(y|\x) \Big] \frac{b_0 p(\rho_n | \bm{\y}, \X)}{p(\x | \bm{\y}, \X)} \\
& = \sum_{\rho_n} \Big[ \frac{\alpha \pi_n}{b} f_{0,\x}(\x) f_{0,y}(y|\x)
+ \frac{1}{b} \sum\limits_{j=1}^K n_j \pi_n f_{j,\x}(\x) f_{j,y}(y|\x) \Big] p(\rho_n | \bm{\y}, \X) \\
& = \sum_{\rho_n} f(y | \bm{\y}, \X,\x, \rho_n) p(\rho_n | \bm{\y}, \X),
\end{align*}
where 
$p(\rho_n | \bm{\y}, \X)$ is the posterior of the clustering $\rho_n$ on $n$ observations, 
\begin{align*}
& b = p(\x | \bm{\y}, \X),\ f_{0,y}(y|\x) = \int \N(y|\x,\bm{\theta}_y) G_0(\bm{\theta}_y),\text{ and }f_{j,y}(y|\x) = \int \N(y|\x,\bm{\theta}_y) p(\bm{\theta}_y | \bm{\y}_j^*, \X_j^*) d \bm{\theta}_y.
\end{align*}
Thus, given each partition (clustering) of the data, the conditional posterior predictive density is a weighted
average of the conditional predictive density with parameters drawn from
baseline distribution and the cluster-wise conditional predictive density. In
practice, the conditional predictive density can be approximated by
averaging over $S$ MCMC posterior samples of $\rho_n$, using: 
\begin{align*}
f(y | \bm{\y}, \X, \x) \approx \frac{1}{S} \sum\limits_{s=1}^S 
\hat{f}^{(s)}(y | \bm{\y}, \X, \x, \rho_n^{(s)}).
\end{align*}

Based on the same covariate-dependent urn scheme structure, the posterior predictive expectation,
conditionally on a new $\x$, is given by:
\begin{align*}
E(Y|\bm{\y}, \X,\x)=\sum_{\rho _{n}}\Big[\frac{\alpha \pi _{n}}{b}f_{0,\x}(\x)E_{0}(Y|\x)+\frac{1}{b}\sum\limits_{j=1}^{K} n_{j}\pi _{n}f_{j,\x}(\x)E_{j}(Y|\x)\Big]p(\rho _{n}|\bm{\y}, \X),
\end{align*}%
where $E_{0}(Y|\x)$ is the expectation of $y$ given $\x$ with distribution 
$f_{0,y}(y|\x)$ and $E_{j}(Y|\x)$ is the expectation of $Y$ given $\x$ with
distribution $f_{j,y}(y|\x)$. The predictive expectation can be
approximated by averaging over MCMC posterior samples of $\rho _{n}$,
i.e., 
\begin{align*}
E(Y|\bm{\y}, \X,\x)\approx \frac{1}{S}\sum\limits_{s=1}^{S}\hat{E}(Y|\x,\bm{\theta}_{1:n}^{(s)}),
\end{align*}%
where:
\begin{align*}
E(Y|\x,\bm{\theta}_{1:n})=\frac{1}{b}\Big[\alpha \int (\mu +\x^{T}\bm{\beta})\prod\limits_{l=1}^{p}\N(x_{l}|m_{l},\tau _{l})G_{0}(d\bm{\theta})+\sum\limits_{j=1}^{n} (\mu _{d_i} +\x^{T}\bm{\beta}_{d_i})\prod\limits_{l=1}^{p}\N(x_{l}|m_{d_i,l},\tau _{d_i,l})\Big],
\end{align*}%
and $b=\alpha \int \prod\limits_{l=1}^{p}\N(x_{l}|m_{l},\tau
_{l})G_{0}(d\bm{\theta})+\sum\limits_{i=1}^{n} \prod\limits_{l=1}^{p}\N(x_{l}|m_{d_i,l},\tau _{d_i,l})$.

\subsection{Variable Selection}
\label{variable selection} The regression coefficients $\beta $'s do not
have positive probabilities of taking on a value of zero in the prior or the
posterior, due to the absolute continuity of the Horseshoe or Normal-Gamma shrinkage prior. However,
it is possible to adopt the Scaled Neighborhood (SN) criterion of \cite{li2010bayesian} for variable selection. 
Specifically, for each data point and its estimated posterior cluster membership $\hat{d}_i$ for $i=1,\dots,n$ 
obtained by an optimal clustering rule (see Section \ref{clustering}), we obtain the MCMC posterior 
samples of coefficients matrix $\bm{B}_{\hat{d}_i} = (\bm{\beta}_{\hat{d}_i,1},\dots,\bm{\beta}_{\hat{d}_i,p})$, 
where $\bm{\beta}_{\hat{d}_i,l}$ (for $l=1,\dots,p$) are $S$-dimensional vectors consisting of posterior samples. 
Then, for each $l=1,\dots,p$, we compute the coordinate-wise (SN) posterior probability $P_{il}$ 
that $\beta _{l}$ is within the scaled neighborhood 
$[-\sqrt{\text{var}(\beta _{\hat{d}_i,l}|\bm{\y}, \X)},\sqrt{\text{var}(\beta _{\hat{d}_i,l}|\bm{\y}, \X)}]$, 
based on marginal posterior variances estimated from the MCMC sampling algorithm. 
Then the decision of whether to exclude the $l$th covariate for observation $i$ in cluster $\hat{d}_i$ depends 
on whether the SN probability $P_{il}$ exceeds a threshold $p^{\ast }$, usually chosen as $p^{\ast }=\frac{1}{2}$.

\subsection{Clustering}
\label{clustering} 
In Bayesian inference, an optimal point estimator $\hat{\rho}$ of the
clustering is obtained by the minimizing solution:
\begin{align*}
\hat{\rho} = \argmin{\rho^{\prime}} \sum \limits_{\rho_n} \mathcal{L}(\rho^{\prime}, \rho_n) p(\rho_n | \bm{\y}, \X), 
\end{align*}%
where $\mathcal{L}$ is a chosen loss function. Here, we 
choose the loss function $\mathcal{L}(a,z)$ by the variation of 
information (VI) \citep{meilua2007comparing}, which is based on information theory 
and is invariant to label-switching of the cluster assignments $\rho_n =\{d_{i}\}_{i=1}^{n}$. 
The VI between two clusterings $(\rho ^{\prime },\rho_n )$ is the sum of their 
Shannon entropies minus twice the information they share. For clustering 
estimation we implemented the greedy algorithm of \cite{rastelli2018optimal}, which aims to find the minimizing solution 
$\hat{\rho}=\argmin{\rho^{\prime}} \frac{1}{S}\sum\limits_{s=1}^{S}\mathcal{L}(\rho ^{\prime},\rho _{n,s})$, 
based on $S$ MCMC posterior samples of the clusterings 
$\rho _{n,s}=\{d_{i,s}\}_{i=1}^{n} \sim  \pi (\rho_n | \bm{\y}, \X)$ for $s=1,\dots ,S$.

According to the random partition characteristic of the joint DPM model, under moderate or high number of covariates $p$, the likelihood for covariates tends to dominate the posterior of the clustering. This could lead to a clustering that is mainly determined by covariate information, esulting in more clusters with only few observations within each cluster when the true covariate distribution is closer to the uniform distribution on a cube \citep{wade2014improving}. To alleviate this potential issue and improve clustering estimation, we specify the prior parameter of DP mass parameter $\alpha$ as $\alpha_{\alpha} = 2$ and $\theta_{\alpha} = 20$ for the real data set analyzed in Sections \ref{realdata}, in order to enforce that the prior expected number of clusters conditional on $\alpha$ is given by $E[\text{number of distinct } d_i's] = \sum\limits_{i=1}^n \frac{\alpha}{\alpha+i-1}$ \citep{escobar1994estimating}, so that smaller $\alpha$ corresponds to a smaller number of clusters on average.

\section{Simulation Study}
\label{simulation}
We compare the HS-DPM model, NG-DPM model, with N-DPM model as our benchmark model, in terms of prediction, variable selection, and clustering accuracy,
over 10 dataset replications of various data simulation conditions, differing by sample size ($n$ = 100, 200, or 400), covariate dimensionality ($p$ = 10, 50, 100, 200, or 300), and number of components ($J$ = 4 or 10). The N-DPM model is the standard DPM model assigned a normal baseline prior distribution which enforces non-shrinkage variable selection, defined by:
\begin{align*}
g_{0}(\mu _{j},\bm{\beta}_{j},\m_{j},\bm{\tau}_{j})=\N_{p+1}\Big((\mu _{j},%
\bm{\beta}_{j})|\bm{\eta},\bm{\Sigma}\Big)\prod\limits_{l=1}^{p}\Big[%
\N(m_{jl}|m_{0},\frac{\tau _{jl}}{n_{0}})\mathit{IG}(\tau _{jl}|\frac{%
	\nu _{0}}{2},\frac{2}{\nu _{0}s_{0}^{2}})\Big],
\end{align*}%
with hyperpriors $\bm{\eta}\sim \N_{p+1}(\bm{0}, 100 \cdot \mathbf{I}_{p+1}))$
and $\bm{\Sigma}^{-1} \sim \text{Wishart}(p+1, 10 \cdot \mathbf{I}_{p+1}))$. Throughout the simulations, all models are assumed the same hyperparameter prior specifications as $n_{0}=0.1$, $m_{0}=0$, $\nu _{0}=2$, $s_{0}^{2}=2$, $\alpha _{0}=2$, $\theta _{0}=2$, and $\alpha _{\alpha }=2$, $\theta _{\alpha }=2$. Data sets were simulated based on a mixture of $J$-component normal mixture for $(y,\x)$, 
with equal mixture weights $1/J$, as follows. Each data set was simulated by sampling 
each data point $(y_{i},\x_{i})$ from $d_{i}\sim \textit{DiscreteUniform}(1,J)$, $x_{il}\sim \N(m_{d_{i},l},\tau _{d_{i},l})$ 
for $l=1,\dots ,p$, and $y_{i}\sim \N(\mu _{d_{i}}+\x_{i}^{T}\bm{\beta}_{d_{i}},\sigma ^{2})$ for $i=1,\dots ,n$. 
For $j=1,\dots,J$, $\bm{\tau}_{j}\equiv (1,\dots ,1)^{T}$, $\m_{j}=j\times (2,\dots
,2)^{T}$, $\mu_{j} = [10-2\times (j-1)] \bm{1}(j \leq 5) + [10-2j] \bm{1}(j>5)$, $\bm{\beta}_{j}=(\underbrace{3,\cdots ,3}_{6-j},0,\cdots ,0)^{T}$ if $j\leq 5$ else $\bm{\beta}_{j}=(\underbrace{-3,\cdots ,-3}_{j-5},0,\cdots ,0)^{T}$, with error variance $\sigma^2=1$. For each simulated data set analyzed, the models are fitted using 5,000 MCMC sampling iterations, which reliably produced samples that converged to the posterior distribution according to
univariate trace plots, after excluding the 2,000 initial samples as burn-in.

We compared the HS-DPM, NG-DPM, and N-DPM models according to the following criteria:
\begin{itemize}
	\item \textbf{Prediction accuracy}: This was measured by $L1=\frac{1}{n_t} \sum\limits_{i=1}^{n_t} |y_{n+i}-E[Y_{n+i}|\x_{n+i}]|$ and $L2=\frac{1}{n_t} \sum\limits_{i=1}^{n_t} (y_{n+i}-E[Y_{n+i}|\x_{n+i}])^2$ predictive error on stand-alone test data of size $n_t=100$, simulated as above.
	
	\item \textbf{Variable selection accuracy}: This is evaluated using Average Area Under the Curve scores A-AUC=$\frac{1}{n}\sum_{i=1}^n \text{AUC}_i$ based on Receiver Operation Characteristic curve analysis. Here $\text{AUC}_i$ is computed according to observation $i$'s corresponding coefficient posterior probability $P_{il}$ (the SN probability; see Section \ref{variable selection}) against its true relevant predictor label $\bm{1}(\beta_{d_i,l} \neq 0)$ for $l=1,\dots,p$, where $\bm{\beta}_{d_i}$ is the true regression coefficient in the cluster group $d_i$ that sampled data point $(\y_i,\x_i)$ belongs to. 
	The parameter estimation performances are also measured by Average Squared Error $\text{ASE} = \frac{1}{n} \sum\limits_{i=1}^n \frac{1}{p} || \hat{\bm{\beta}}_{\hat{d}_i}^{\text{med}} - \bm{\beta}_{d_i} ||^2$,
	where $\hat{d}_i$ is the $i$'s cluster membership index from estimated optimal clustering rule $\hat{\rho}=\{\hat{d}_1,\dots,\hat{d}_n\}$ (Section \ref{clustering}), and $\hat{\bm{\beta}}_{\hat{d}_i}^{\text{med}}$ is the posterior median of slope
	coefficients based on the estimated cluster group $\hat{d}_i$.
	
	\item \textbf{Clustering accuracy}: Clustering performance is evaluated by the the Adjusted Rand Index \citep{hubert1985comparing}, and the estimated number of clusters, $\hat{J}$, obtained from the method in Section \ref{clustering}.
\end{itemize}

The results of the simulation study are as follows. As shown in Table \ref{tab:compare_1}, the HS-DPM and NG-DPM models generally outperformed the N-DPM in terms of predictive criteria and coefficient estimation under most of the conditions defined by $n$, $p$ and $J$. The HS-DPM model emerged as the winner in most of the simulation scenarios. For example, under the setting of $n=200$ and $J=4$, for low dimensional cases ($p=10$ or 50), all three DPM models perform comparatively in prediction, coefficient estimation and variable selection accuracy. While under moderate to high dimensional scenarios ($p=200$ or 300), the HS-DPM and NG-DPM models had significantly better predictive and variable selection performances according to the $L1$, $L2$, ASE and A-AUC score, thanks to the effect of Horseshoe or Normal-Gamma prior which can provide better coefficient estimations by adaptively shrinking the coefficients of irrelevant covariates towards zero. In terms of clustering accuracy, all three models perform competitively well due to the nature of joint modeling on the covariates and response, which is usually dominated by the similarity among covariate values when $p$ is not small.


\section{Real Data Illustration}
\label{realdata}
We also illustrate our proposed HS-DPM and NG-DPM models through the analysis of the Tehran Residential Building data set \citep{rafiei2016novel}, obtained from the UCI Machine Learning repository. The data set contains 372 single-family residential buildings in Tehran, Iran, during 1993 through 2008, each building having between 3 to 9 stories. The data set contains corresponding observations of 9 project physical and financial variables, including total floor area (V1), lot area (V2), total preliminary estimated construction cost (V3), preliminary estimated construction cost (V4), equivalent preliminary estimated construction cost in base year (V5), duration of construction (V6), unit price at project beginning (V7), sales price, and construction cost. The data set also includes corresponding observations of 19 economic variables in 5 time-lag numbers before the initial construction date of the building; and observations of the variable named profit, defined by the difference of sales price and construction cost of the building project.

\begin{figure}[hbt!]
	\centering
	\includegraphics[width=\textwidth]{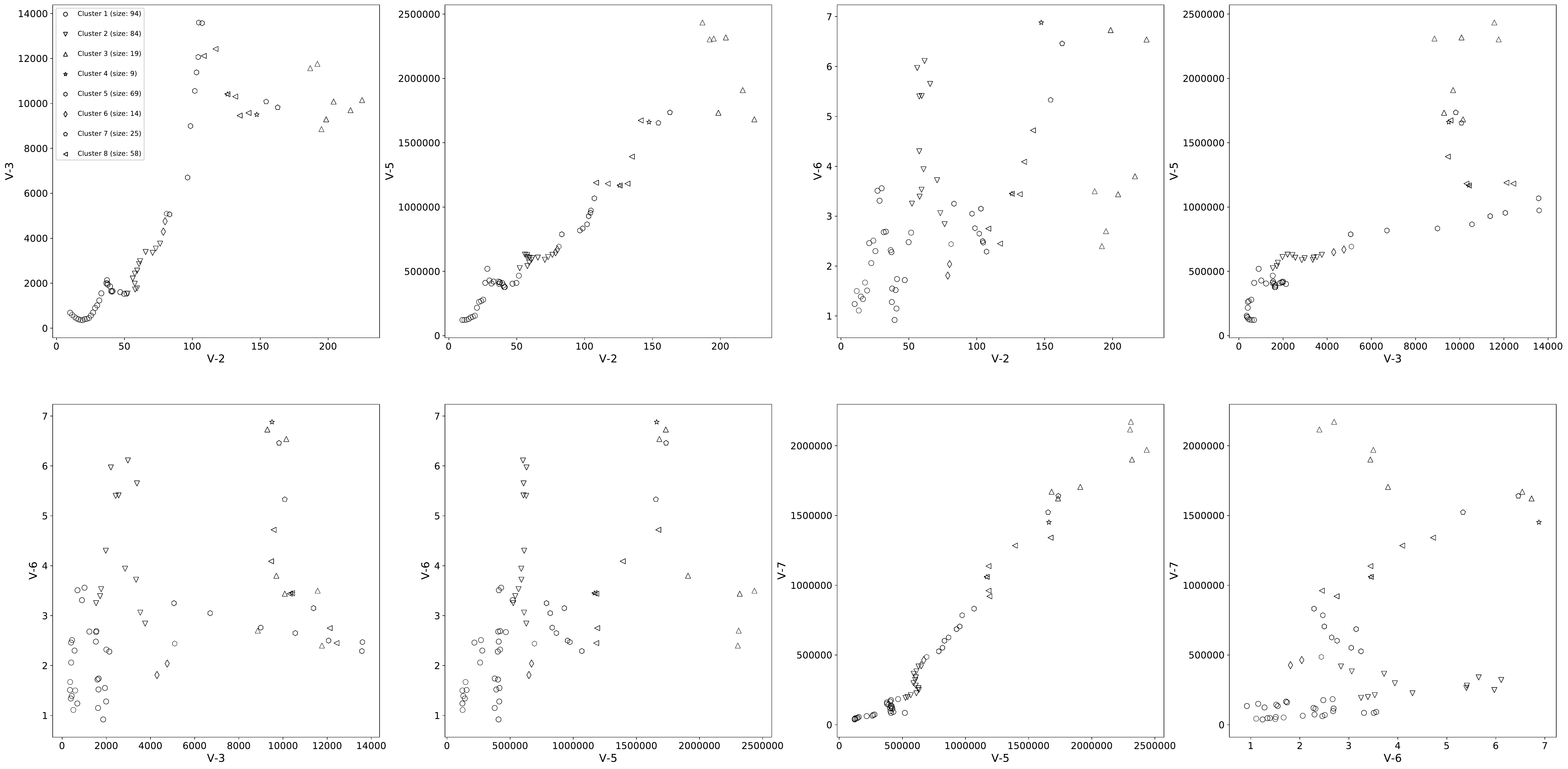}
	\caption{For the HS-DPM model, the clustering estimate, according to 2-dimensional views of covariate observations.}
	\label{fig:residential_hs}
\end{figure}

The aim of the data analysis was to predict log profit of the construction project as a function of 102 covariates, including the physical or financial variables (V1 to V7) and all the 95 economic variables (7 + 19(5) = 102). Also, the aim was to estimate the latent clustering groups of the building projects, and to identify the subsets of relevant (and irrelevant) covariates of profit, for each cluster.

\begin{figure}[hbt!]
	\centering
	\includegraphics[width=0.9\textwidth]{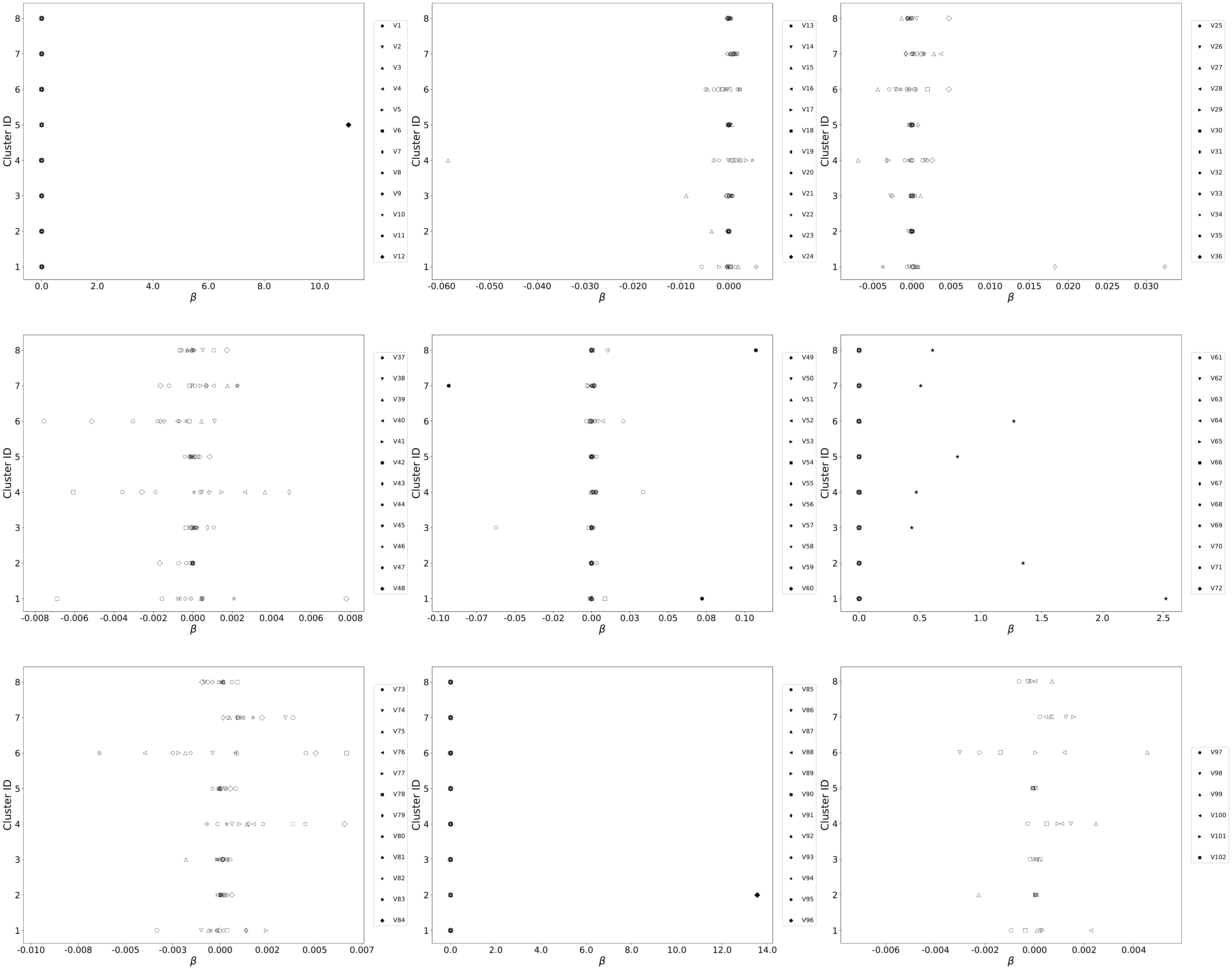}
	\caption{For the HS-DPM model, the marginal posterior medians of all 102 regression coefficients, and cluster-wise variable selection results. A filled marker indicates a significant covariate according to the SN criterion using threshold $p^* = 0.5$.}
	\label{fig:realdata_boxplot}
\end{figure}

The HS-DPM, NG-DPM, and the benchmark N-DPM models were each fitted to the residential data set, using the same prior distribution and MCMC algorithm specifications used in Section \ref{simulation}. As an exception, the precision parameter $\alpha$ was assigned a $\mathit{Ga}(2, 20)$ prior distribution. As another benchmark model, we also analyzed the data using the Bayesian Horseshoe (HS) normal linear regression model from R code with default prior specifications in \cite{gramacy2019package}, while this model does not perform clustering. The predictive performances of all four models were also evaluated by the mean of $L1$ and $L2$ prediction error, measured through 5-fold cross validation, based on a train-test split ratio of 0.2. For each fold of splitting, we normalized the training data observations of each variable into $z$-scores having mean 0 and variance 1, and then fit the transformation on the test data accordingly.


Both the HS-DPM and NG-DPM models significantly outperformed the benchmark N-DPM model in terms of predictive accuracy, apparently because of more accurate coefficient estimation under the cluster-wise "high-dimensional" scenarios. The (mean $L2$) prediction errors for the models were, respectively, HS-DPM (0.08), NG-DPM (0.27), HS (0.28), and N-DPM (0.89); while the four models were similarly ordered with respect to mean $L1$ prediction error. For the predictive comparison between our proposed models and HS, the HS-DPM model was still the obvious winner, while HS and NG-DPM performed comparatively.

In addition, we inspected the respective clustering estimates of the HS-DPM, NG-DPM, and N-DPM models, obtained from the entire data set.
Clustering solutions are similar for three models, each of which estimated 8 clusters from the data, due to the same prior specification on the mass parameter and the joint DPM clustering rule. The estimated clustering for the HS-DPM model, along with corresponding cluster size, is presented in Figure \ref{fig:residential_hs}, which lends to useful interpretations. For example, as the largest group, Cluster 1 generally displays construction projects with small lot area and low construction cost, and is one of the largest cluster groups. In contrast, the building projects allocated into the third cluster group tend to have larger lot area and higher construction costs.

Figure \ref{fig:realdata_boxplot} presents the cluster-wise variable selection results for the HS-DPM model, indicating which variables were relevant or irrelevant in the prediction of log profit. While most of the regression coefficients estimates were similar across different clusters, some coefficient estimates and sparsity patterns differed across the cluster groups, with V70 (total floor areas of building permits issued by city in 4th time lag) as a relevant covariate for prediction for all cluster groups.

\section{Conclusions}
We proposed two novel joint DPM models, namely the HS-DPM and the NG-DPM, each of which adopt a continuous shrinkage (baseline) prior on the regression coefficients, to provide flexible prediction and cluster-wise variable selection. The development of these models was motivated by the fact that most of the existing literature either focuses on the standard benchmark DPM model, which adopts a conjugate non-shrinkage normal baseline prior on regression coefficients which allows for easier posterior computations, but does not allow for heterogeneous variable selections across different cluster groups; or focuses on the spike-and-slab baseline prior for covariate selection based on latent binary variables, which is computationally costly when the number of covariates is large.

Our models were able to provide inference on variable selection, while maintaining a  computational cost comparable with that of the standard benchmark DPM model. In the simulation study, we have shown that the HS-DPM and the NG-DPM models generally provided better prediction, coefficient estimation, and cluster-wise variable selection accuracy than the benchmark model, especially when number of covariates is much larger than the within-cluster sample size. We also highlighted the advantage in prediction for our proposed models, and presented corresponding clustering and cluster-wise variable selection idea through a real data application. The improvement in prediction compared with the Bayesian Horseshoe normal linear model was also presented to show the benefits of allowing for heterogeneous sparsity patterns across different clusters of the data set.

For future research, the novel models can be extended to mixtures of generalized linear models in order to address additional types of dependent response variables, such as binary response. The models can also be extended to other types nonparametric priors which generalize the DP to offer more flexibility in the analysis of the relationship between covariates and response.

\section*{Acknowledgments}
The article represents work from the first author's Ph.D. dissertation, and is supported by National Science Foundation grant SES-1156372 awarded to the second author.

\bibliography{reference}

%
%

\begin{table}[b]
	\centering
	\begin{adjustbox}{width=0.8\textwidth,center=\textwidth}
		\begin{tabular}{cccccclccc}
			\multicolumn{3}{c}{Condition} & \multicolumn{3}{c}{$L1$}                                                                                              &  & \multicolumn{3}{c}{$L2$}                                                                                               \\
			$n$      & $p$      & $J$     & HS                                    & NG                                    & N                                     &  & HS                                     & NG                                    & N                                     \\ \cline{1-6} \cline{8-10} 
			100      & 100       & 4 &   \textbf{0.98 (0.10)} & 2.54 (0.39) &13.86 (2.46) &&   \textbf{1.57 (0.33)} & 11.75 (3.77) & 314.31 (111.08)\\ 
			100      & 200       & 4 &   \textbf{0.95 (0.15)} & 3.79 (0.58) &56.25 (7.09) &&   \textbf{1.43 (0.44)} & 23.53 (6.57) & 4962.52 (1108.27)\\
			200      & 10       & 4 &  \textbf{0.96 (0.26)} & \textbf{0.96 (0.25)} & 0.98 (0.22) &&  2.68 (3.91) & 2.65 (3.73) & \textbf{2.54 (3.32)}\\ 
			200      & 50       & 4       &  \textbf{0.93 (0.13)} & 0.96 (0.15)                           & 1.75 (0.25)                           &  &  \textbf{1.35 (0.39)} & 1.46 (0.44)                            &4.91 (1.55)                           \\
			200      & 100      & 4       &  \textbf{0.95 (0.13)} & 1.12 (0.14)                           &2.69 (0.24)                           &  &  \textbf{1.43 (0.32)} & 1.85 (0.44)                            &11.35 (1.76)                          \\
			200      & 200       & 4 &  \textbf{0.93 (0.13)} &1.33 (0.40) & 23.81 (4.78) &&   \textbf{1.34 (0.40)} & 3.67 (3.22) &911.31 (399.75)\\ 
			200      & 300       & 4 &  \textbf{1.27 (0.13)} & 2.02 (0.30) & 64.22 (6.91) &&  \textbf{3.45 (1.02)} & 8.09 (2.62) & 6682.93 (1428.73)\\ 			
			200      & 50       & 10      &  \textbf{1.64 (0.48)} & 2.57 (0.91)                           &4.48 (1.77)                           &  &  \textbf{6.89 (4.11)} &18.16 (17.67)                          & 48.01 (45.05)                         \\
			200      & 100      & 10      &  \textbf{1.55 (0.15)} & 3.60 (1.09)                           & 8.13 (2.01)                           &  &  \textbf{4.61 (0.95)} & 28.06 (20.87)                          &113.87 (53.20)                        \\			
			200      & 200       & 10  &  \textbf{2.48 (0.33)} & 5.99 (1.11) &31.52 (6.06) &&   \textbf{11.55 (4.21)} & 67.13 (25.35) &1636.01 (623.85)\\ 
			400      & 50       & 4       &  \textbf{0.80 (0.05)} & 0.83 (0.06)                           &1.08 (0.06)                           &  &  \textbf{1.02 (0.12)}                           & 1.07 (0.11)  & 1.78 (0.20)                           \\
			400      & 100      & 4       & \textbf{0.87 (0.08)} & 0.98 (0.07)                           &  1.76 (0.18)                           &  &  \textbf{1.22 (0.17)} & 1.51 (0.19)                            &4.99 (0.93)                           \\
			400      & 200       & 4 &  \textbf{0.88 (0.13)} & 1.03 (0.13) &  2.75 (0.35) &&   \textbf{1.24 (0.32)} & 1.64 (0.39) &11.52 (2.76)\\ 
			400      & 50       & 10      &  \textbf{0.94 (0.16)} & 1.33 (0.46)                           &3.00 (1.11)                           &  &  \textbf{1.64 (0.94)} & 4.33 (4.12)                            & 17.39 (12.52)                         \\			
			400      & 100      & 10      &  \textbf{1.27 (0.29)} & 1.59 (0.43)                           &4.74 (1.45)                           &  &  \textbf{3.09 (1.67)} & 4.90 (2.68)                            &53.81 (54.54)                         \\			
			\multicolumn{3}{c}{Condition} & \multicolumn{3}{c}{ARI}                                                                                               &  & \multicolumn{3}{c}{$\hat{J}$}                                                                                          \\
			$n$      & $p$      & $J$     & HS                                    & NG                                    & N                                     &  & HS                                     & NG                                    & N                                     \\ \cline{1-6} \cline{8-10} 
			100      & 100       & 4  & \textbf{1.00 (0.00)}                           & \textbf{1.00 (0.00)}                            & \textbf{1.00 (0.00)}                            &  & \textbf{4.00 (0.00)}                            & \textbf{4.00 (0.00)}        & \textbf{4.00 (0.00)} \\ 
			100      & 200       & 4 & \textbf{1.00 (0.00)}                           & \textbf{1.00 (0.00)}                            & \textbf{1.00 (0.00)}    &  & \textbf{4.00 (0.00)}     & \textbf{4.00 (0.00)}        & \textbf{4.00 (0.00)}  \\			
			200      & 10       & 4 & \textbf{1.00 (0.00)}                           & \textbf{1.00 (0.00)}                            & \textbf{1.00 (0.00)}      &  & \textbf{4.00 (0.00)}    & \textbf{4.00 (0.00)}         & \textbf{4.00 (0.00)}         \\ 
			200      & 50       & 4       & \textbf{1.00 (0.00)}                           & \textbf{1.00 (0.00)}                            & \textbf{1.00 (0.00)}      &  & \textbf{4.00 (0.00)}    & \textbf{4.00 (0.00)}         & \textbf{4.00 (0.00)}         \\
			200      & 100      & 4       & \textbf{1.00 (0.00)}                           & \textbf{1.00 (0.00)}                            & \textbf{1.00 (0.00)}      &  & \textbf{4.00 (0.00)}    & \textbf{4.00 (0.00)}         & \textbf{4.00 (0.00)}         \\
			200      & 200       & 4 &  \textbf{1.00 (0.00)}                           & \textbf{1.00 (0.00)}                            & \textbf{1.00 (0.00)}    &  & \textbf{4.00 (0.00)}     & \textbf{4.00 (0.00)}        & \textbf{4.00 (0.00)}  \\			
			200      & 300       & 4 &  \textbf{1.00 (0.00)} & 0.90 (0.06) & 0.96 (0.05) && \textbf{4.00 (0.00)}  & 3.67 (0.47) & 3.86 (0.35)\\ 
			200      & 50       & 10      &  0.92 (0.01)                           & 0.89 (0.02)                           & \textbf{0.93 (0.02)} &  &  9.29 (0.45)                           &8.89 (0.99)                            & \textbf{9.30 (0.78)} \\
			200      & 100      & 10      &  \textbf{0.92 (0.01)} & 0.90 (0.01)                           & 0.86 (0.02)                           &  &  \textbf{9.25 (0.43)} &9.10 (0.70)                            & 8.60 (0.92)                           \\			
			200      & 200       & 10 &  0.78 (0.01) & \textbf{0.80 (0.02)} &0.73 (0.03) &&  7.67 (0.47) & \textbf{8.00 (1.05)} & 7.30 (1.10)\\ 
			400      & 50       & 4       &  \textbf{1.00 (0.00)}                           & \textbf{1.00 (0.00)}                            & \textbf{1.00 (0.00)}    &  & \textbf{4.00 (0.00)}     & \textbf{4.00 (0.00)}        & \textbf{4.00 (0.00)}  \\			
			400      & 100      & 4       &  \textbf{1.00 (0.00)}                           & \textbf{1.00 (0.00)}                            & \textbf{1.00 (0.00)}    &  & \textbf{4.00 (0.00)}     & \textbf{4.00 (0.00)}        & \textbf{4.00 (0.00)}  \\			
			400      & 200       & 4 &  \textbf{1.00 (0.00)}                           & \textbf{1.00 (0.00)}                            & \textbf{1.00 (0.00)}    &  & \textbf{4.00 (0.00)}     & \textbf{4.00 (0.00)}        & \textbf{4.00 (0.00)}  \\			
			400      & 50       & 10      & \textbf{0.98 (0.01)} & 0.88 (0.02)                           & 0.93 (0.01)                           &  &  \textbf{9.75 (0.43)} & 8.86 (0.99)                            &9.30 (0.46)                           \\			
			400      & 100      & 10      & \textbf{0.92 (0.02)} & 0.88 (0.02)                           & 0.89 (0.02)                           &  &  \textbf{9.12 (0.78)} & 8.70 (0.78)                            &8.90 (0.83)                           \\
			\multicolumn{3}{c}{Condition} & \multicolumn{3}{c}{ASE}                                                                                               &  & \multicolumn{3}{c}{A-AUC}                                                                                              \\
			$n$      & $p$      & $J$     & HS                                    & NG                                   & N                                     &  & HS                                     & NG                                    & N                                     \\ \cline{1-6} \cline{8-10} 
			100      & 100       & 4 &  \textbf{0.00 (0.00)} & 0.13 (0.05) & 3.33 (1.12) &&  \textbf{1.00 (0.00)} &  0.99 (0.01) &0.76 (0.13)\\ 
			100      & 200       & 4 &   \textbf{0.00 (0.00)} & 0.12 (0.01) &26.39 (2.54) &&   \textbf{1.00 (0.00)} & 0.94 (0.04) &0.49 (0.20)\\
			200      & 10       & 4 &    \textbf{0.02 (0.00)} & \textbf{0.02 (0.01)} & \textbf{0.02 (0.01)} &&  \textbf{1.00 (0.00)} & \textbf{1.00 (0.00)} & \textbf{1.00 (0.00)}\\ 
			200      & 50       & 4       &  \textbf{0.00 (0.00)} & 0.01 (0.00)                           & 0.08 (0.02)                           &  &  0.88 (0.03)                           &0.90 (0.03)                            & \textbf{0.92 (0.04)} \\
			200      & 100      & 4       & \textbf{0.00 (0.00)} & \textbf{0.00 (0.00)} & 0.10 (0.01)                           &  & 0.89 (0.04)                            & 0.89 (0.03)                           & \textbf{0.99 (0.01)} \\
			200      & 200       & 4 &   \textbf{0.00 (0.00)} & 0.01 (0.01) &4.43 (1.72) &&  \textbf{1.00 (0.00)} & \textbf{1.00 (0.00)} & 0.73 (0.14)\\ 
			200      & 300       & 4 & \textbf{0.00 (0.00)} & 0.02 (0.01) & 24.53 (3.18) && \textbf{1.00 (0.00)}  & \textbf{1.00 (0.00)} & 0.42 (0.12)\\ 
			200      & 50       & 10      &  \textbf{0.09 (0.06)} &0.22 (0.17)                           & 0.84 (0.92)                           &  &  \textbf{0.93 (0.02)} & 0.91 (0.05)                            &0.85 (0.13)                           \\
			200      & 100      & 10      & \textbf{0.04 (0.02)} & 0.17 (0.07)                           & 0.77 (0.24)                           &  & 0.89 (0.04)                            & \textbf{0.92 (0.02)} & 0.53 (0.12)                           \\
			200      & 200       & 10 &   \textbf{0.05 (0.00)} & 0.14 (0.01) &8.15 (2.22) &&   \textbf{0.96 (0.02)} & 0.67 (0.14) & 0.50 (0.14)\\ 
			400      & 50       & 4       & \textbf{0.00 (0.00)} & \textbf{0.00 (0.00)} & 0.02 (0.00)                           &  & 0.89 (0.04)                            & 0.89 (0.03)                           & \textbf{0.90 (0.03)} \\
			400      & 100      & 4       & \textbf{0.00 (0.00)} & \textbf{0.00 (0.00)} & 0.04 (0.00)                           &  & 0.90 (0.05)                            & 0.90 (0.04)                           & \textbf{0.96 (0.03)} \\
			400      & 200       & 4 &  \textbf{0.00 (0.00)} & \textbf{0.00 (0.00)} & 0.05 (0.01) &&  \textbf{1.00 (0.00)} & \textbf{1.00 (0.00)} & \textbf{1.00 (0.00)}\\ 
			400      & 50       & 10      &  \textbf{0.02 (0.02)} & 0.05 (0.03)                           & 0.26 (0.16)                           &  &  \textbf{0.94 (0.02)} & 0.92 (0.02)                            &\textbf{0.94 (0.03)} \\			
			400      & 100      & 10      & \textbf{0.02 (0.02)} & 0.04 (0.02)                           & 0.38 (0.34)                           &  & \textbf{0.94 (0.02)}  & \textbf{0.94 (0.02)} & 0.93 (0.04) \\			
			\cline{1-6} \cline{8-10}                       
		\end{tabular}
	\end{adjustbox}
	\caption{For the HS-DPM, NG-DPM, and N-DPM models, mean (standard error) of predictive accuracy, variable selection accuracy, and clustering accuracy statistics, obtained over 10 replications, for each of the simulation conditions. Best performance values are indicated in bold.}
	\label{tab:compare_1}
\end{table}


\appendix

\section{MCMC Algorithm for the HS-DPM Model}
\label{hs_mcmc}
Equation (\ref{model:joint}) gives the joint posterior distribution of the model parameters, up to a proportionality constant. Based on the "scale mixture" relationship between half-Cauchy distribution and Inverse-Gamma distribution \citep{makalic2015simple}, we can rewrite model (\ref{hs_prior}) by augmentation of hyperparameters as:
\begin{subequations}
	\label{hs_prior_1}
	\begin{align}
	& \bm{\Gamma}_j = \text{diag}(\gamma_{j1}^2,\dots,\gamma_{jp}^2) \equiv \text{
		diag}(\bm{\gamma}^2_j), \ j=1,2,\dots  \\
	& \pi(\bm{\gamma}^2_j | \nu_j) = \prod \limits_{l=1}^{p} \mathit{IG}(\gamma_{jl}^2 | 1/2,1 / \nu_{jl}), \ j=1,2,\dots \\
	& \pi(\zeta^2_j | \xi_j)  = \mathit{IG}(\zeta_j^2 | 1/2, 1 / \xi_j), \ j=1,2,\dots \\
	& \pi(\nu_{1j},\dots,\nu_{pj},\xi_j) = \prod\limits_{l=1}^p \mathit{IG}(\nu_{jl} | 1/2,1) \mathit{IG}(\xi_j | 1/2,1), \ j=1,2,\dots.
	\end{align}
\end{subequations}
Then the MCMC algorithm for sampling the posterior distributions from HS-DPM model is described as follows:

\textbf{Step 0: Initialization:} Denote $s$ as the iteration index within
MCMC algorithm. Initialize with starting values by setting $s=0$, draw $d_{i}^{(0)}\sim \mathit{DiscreteUniform}(1,n)$ for $i=1,\cdots ,n$, and ${\sigma ^{2}}^{(0)} \sim \mathit{IG}(\alpha _{0},\theta_{0}) $. Then for $j=1,\cdots ,M^{(0)}=\max\limits_{i}d_{i}^{(0)}$, $\nu_{jl}^{(0)} \sim \mathit{IG}(1/2,1)$, $\xi_j^{(0)} \sim \mathit{IG}(1/2,1)$, ${\gamma_{jl}^2}^{(0)} \sim \mathit{IG}(1/2,1 / \nu_{jl}^{(0)})$, ${\zeta^2_j}^{(0)} \sim \mathit{IG}(1/2,1 / \xi_{j}^{(0)})$. We also initialize starting values for $\mu _{j}^{(0)} \sim \N(0,\nu_{\mu })$, $\bm{\beta}_{j}^{(0)} \sim \N_{p}(\bm{0}_{p}, {\zeta^2_j}^{(0)}{\sigma ^{2}}^{(0)} \bm{\Gamma}_j^{(0)} )$, $m_{jl}^{(0)}\sim \N(m_{0},\tau _{jl}^{(0)}/n_{0})$, $\tau_{jl}^{(0)}\sim \mathit{IG}(\frac{\nu _{0}}{2},\frac{2}{\nu
	_{0}s_{0}^{2}})$, and mass parameter $\alpha ^{(0)}\sim \mathit{Ga}(\alpha _{\alpha
},\theta _{\alpha })$.

Then, for each iteration $s=1,\cdots ,S$, draw from the full conditional 
posterior distributions described in the following steps:

\textbf{Step 1: Draw mixture weights} $w_{j}^{(s)}$: For $j=1,\cdots ,M^{(s)}=\max\limits_{i}d_{i}^{(s-1)}$, take $%
n_j^{(s)}=\sum\limits_{i=1}^{n}\bm{1}(d_{i}^{(s-1)}=j)$, $%
m_{j}^{(s)}=\sum\limits_{i=1}^{n}\bm{1}(d_{i}^{(s-1)}>j)$. Then for $%
j=1,\cdots ,M^{(s)}$, draw $v_{j}^{(s)}\sim \mathit{Beta}(1+n_j^{(s)},\alpha
^{(s-1)}+m_{j}^{(s)})$, let $w_{j}^{(s)}=v_{j}^{(s)}\prod%
\limits_{l<j}(1-v_{j}^{(s)})$, and draw $u_{i}^{(s)}\sim \mathit{Unif}(0,w_{d_{i}^{(s-1)}}^{(s)})$ for $i=1,\cdots ,n$, where $\mathit{Unif}(c,d)$ is the Uniform distribution with parameters $c$ and $d$. For $j=M^{(s)}+1,\cdots
,N^{(s)}$, draw $v_{j}^{(s)}\sim \mathit{Beta}(1+n_j^{(s)},\alpha
^{(s-1)}+m_{j}^{(s)})$ until the smallest $N^{(s)}$ is obtained such that $%
\sum\limits_{j=1}^{N^{(s)}}w_{j}^{(s)}>\max\limits_{i}(1-u_{i}^{(s)})$.%

\textbf{Step 2}: For $j=1,\cdots ,M^{(s)}$, update $\bm{\nu}_j^{(s)}$, $\xi_j^{(s)}$, ${\bm{\gamma}^2_{j}}^{(s)}$, ${\zeta^2_{j}}^{(s)}$, $\mu _{j}^{(s)},\bm{\beta}_{j}^{(s)}$, $\m_{j}^{(s)},\bm{\tau}_{j}^{(s)}$:

\textbf{2.1}: For $l=1,\dots,p$, draw from:
\begin{align*}
& \nu_{jl}^{(s)} | {\gamma^2_{jl}}^{(s-1)} \propto \pi(\nu_{jl}) \pi({\gamma^2_{jl}}^{(s-1)} | \nu_{jl}) \sim \mathit{IG}(1, 1 + \frac{1}{{\gamma_{jl}^2}^{(s-1)}}).
\end{align*}

\textbf{2.2}: Draw from:
\begin{align*}
&\xi_j^{(s)} | {\zeta_j^2}^{(s-1)} \propto \pi(\xi_j) \pi({\zeta_j^2}^{(s-1)} | \xi_j) \sim \mathit{IG}(1, 1 + \frac{1}{{\zeta_j^2}^{(s-1)}}).
\end{align*}

\textbf{2.3}: For $l=1,\dots,p$, draw from:
\begin{align*}
& {\gamma_{jl}^2}^{(s)} | \nu_{jl}^{(s)}, {\zeta_j^2}^{(s-1)}, \beta_{jl}^{(s-1)}, {\sigma^2}^{(s-1)} \propto \pi(\gamma_{jl}^2 | \nu_{jl}^{(s)}) \pi(\beta_{jl}^{(s-1)} | \gamma_{jl}^2) \sim \mathit{IG}(1, \frac{1}{\nu_{jl}^{(s)}} + \frac{{\beta_{jl}^{(s-1)}}^2}{2  {\zeta_j^2}^{(s-1)} {\sigma^2}^{(s-1)}}).
\end{align*}

\textbf{2.4}: Draw from:
\begin{align*}
& {\zeta_j^2}^{(s)} |  \xi_j^{(s)},  {\gamma_{jl}^2}^{(s)}, \beta_{jl}^{(s-1)}, {\sigma^2}^{(s-1)} \propto \pi(\zeta^2_j |  \xi_j^{(s)}) \pi( {\bm{\beta}}^{(s-1)} | \zeta_j^2) \sim \mathit{IG}( \frac{p+1}{2}, \frac{1}{\xi_j^{(s)}} +  \sum\limits_{l=1}^{p} \frac{{\beta_{jl}^2}^{(s-1)}}{{\gamma_{jl}^2}^{(s)}} ).
\end{align*}

\textbf{2.5}: Draw from:
\begin{align*}
&& \mu _{j}^{(s)},\bm{\beta}_{j}^{(s)} | {\bm{\gamma}_{jl}^2}^{(s)}, {\zeta_j^2}^{(s)}, {\sigma^2}^{(s-1)} &\propto \prod\limits_{d_{i}^{(s-1)}=j}\N(y_i|\x_i,\mu _{j},\bm{\beta}_{j},{\sigma ^{2}}^{(s-1)}) \pi(\bm{\beta}_j | {\bm{\gamma}_{jl}^2}^{(s)}, {\zeta_j^2}^{(s)}, {\sigma^2}^{(s-1)}) \pi(\mu_j)\\
&&& \sim \N_{p+1}(B^{-1}{\tilde{\X}_j}^{*^T} \bm{\y}_j^*,{\sigma ^{2}}^{(s-1)}B^{-1}),
\end{align*}
where $B={\tilde{\X}_j}^{*^T} {\tilde{\X}_j}^{*} +{\sigma^{2}}^{(s-1)} {\bm{\Gamma}_{j}^*}^{(s)}$, ${\tilde{\X}_j}^{*^T} =[\bm{1}_{n_j},\X_j^*]$, ${\bm{\Gamma}_{j}^*}^{(s)} = \frac{1}{{\zeta_j^2}^{(s)}} \text{diag}\Big(0,1 / {\gamma_{j1}^2}^{(s)},\cdots ,1 / {\gamma_{jp}^2}^{(s)}\Big)$.

\textbf{2.6}: For $l=1,\dots ,p$, draw from:
\begin{align*}
\tau _{jl}^{(s)}\propto \prod\limits_{d_{i}^{(s-1)}=j} \N(x_{il}|m_{jl}^{(s-1)},\tau _{jl})\pi (\tau _{jl})\sim \mathit{IG}(\frac{%
	\nu ^{\ast }}{2},\frac{2}{{s^{\ast }}^{2}\nu ^{\ast }}),
\end{align*}%
where $n^{\ast }=n_{0}+n_{j}$, $\nu ^{\ast }=\nu _{0}+n_{j}$, $\bar{x}_{n_{j},l}=\frac{1}{n_{j}}\sum\limits_{i=1}^{n_{j}}x_{il}$, and ${%
	s^{\ast }}^{2}=\frac{1}{\nu ^{\ast }}\Big[\sum\limits_{i=1}^{n_{j}}(x_{il}-\bar{x}_{n_{j},l})^{2}+s_{0}^{2}\nu _{0}+\frac{n_{0}n_{j}}{%
	n^{\ast }}(\bar{x}_{n_{j},l}-m_{0})^{2}\Big]$.

\textbf{2.7}: For $l=1,\dots ,p$, draw from:
\begin{align*}
m_{jl}^{(s)}|\tau _{jl}^{(s)}\propto \prod\limits_{d_{i}^{(s-1)}=j} \N(x_{il}|m_{jl},\tau _{jl}^{(s)})\pi (m_{jl}|\tau _{jl}^{(s)})\sim \N(m^{\ast
},\frac{\tau _{jl}^{(s)}}{n^{\ast }}),
\end{align*}%
where $m^{\ast }=(n_{j}\bar{x}_{n_{j},l}+n_{0}m_{0}) / (n_{j}+n_{0})$.

If $N^{(s)}>M^{(s)}$, then for $j=M^{(s)}+1,\cdots ,N^{(s)}$, draw $\nu_{jl}^{(s)} \sim \mathit{IG}(1/2,1)$, $\xi_j^{(s)} \sim \mathit{IG}(1/2,1)$, ${\gamma_{jl}^2}^{(s)} \sim \mathit{IG}(1/2,1 / \nu_{jl}^{(s)})$, ${\zeta^2_j}^{(s)} \sim \mathit{IG}(1/2,1 / \xi_{j}^{(s)})$, $\mu _{j}^{(s)} \sim \N(0,\nu_{\mu })$, $\bm{\beta}_{j}^{(s)} \sim \N_{p}(\bm{0}_{p}, {\zeta^2_j}^{(s)}{\sigma ^{2}}^{(s-1)} \bm{\Gamma}_j^{(s)} )$, $m_{jl}^{(s)} \sim \N(m_{0},\tau _{jl}^{(s)}/n_{0})$, $\tau_{jl}^{(s)}\sim \mathit{IG}(\frac{\nu _{0}}{2},\frac{2}{\nu_{0}s_{0}^{2}})$, 

\textbf{Step 3}: For $i=1,\cdots,n$, sample $%
d_i^{(s)} = j$ with probability proportional to:
\begin{align*}
\bm{1}(w_j^{(s)}>u_i^{(s)}) \N(y_i | \mu_j^{(s)}+\x_i^T\bm{\beta}_j^{(s)},{%
	\sigma^2}^{(s-1)}) \N_p(\x_i|\m_j^{(s)}, \bm{\tau}_j^{(s)}) , \text{ for } j=1,\cdots,N^{(s)}.
\end{align*}

\textbf{Step 4}: Once we obtain $d_i^{(s)}$ for $i=1,\dots,n$, we have the distinct set of coefficients $\{ {\bm{\beta}_1^*}^{(s)},\dots, {\bm{\beta}_K^*}^{(s)} \}$ among $\{ \bm{\beta}_{d_i}^{(s)} \}$, and corresponding distinct sets $\{ {{\gamma^*_1}^2}^{(s)},\dots, {{\gamma^*_K}^2}^{(s)} \}$ and $\{ {{\xi^*_1}^2}^{(s)},\dots, {{\xi^*_K}^2}^{(s)} \}$, where $K$ is number of unique $d_i^{(s)}$'s. Then draw from:
\begin{align*}
&& {\sigma^2}^{(s)} | \bm{\beta}^*_k, \mu^*_k &\propto \prod\limits_{i=1}^n \N(y_i | \mu_{d_i}^{(s)}+\x_i^T\bm{\beta}_{d_i}^{(s)},{\sigma^2}) \pi(\sigma^2) \pi({\bm{\beta}})
\\
&&& \sim \mathit{IG}\Big( (n+p)/2 + \alpha_0, \sum\limits_{i=1}^n(y_i -
\mu_{d_i}^{(s)} - \x_i^T \bm{\beta}_{d_i}^{(s)})^2 / 2 + \sum\limits_{k=1}^K \frac{1}{2{{\xi^*_k}^2}^{(s)}} \sum\limits_{l=1}^p \frac{{{\beta_{lk}^*}^2}^{(s)}}{{{\gamma_{lk}^*}^2}^{(s)}} + \theta_0 \Big).
\end{align*}

\textbf{Step 5}: Update $\alpha^{(s)}$ by drawing from a Gamma distribution
with shape $\alpha_{\alpha} + K - \bm{1}(u > \{O/(1+O)\})$ and scale $\theta_{\alpha}-\log(\eta)$, where  $\eta
\sim \mathit{Beta}(\alpha^{(s-1)}+1, n)$, $u \sim \textit{Unif}(0,1)$ and $O
= (\alpha_{\alpha} + K - 1)/( \{\theta_{\alpha} - \log(\eta)\} n)$ \citep{escobar1995bayesian}.

Sampling updates from Steps 1 through 5 are repeated for a large number of iterations until the MCMC chain has displayed good mixing, according to trace plots.

\section{MCMC Algorithm for the NG-DPM Model}
\label{ng_mcmc}
The MCMC algorithm for sampling the posterior distributions from NG-DPM model is described as follows:

\textbf{Step 0: Initialization:} Denote $s$ as the iteration index within
MCMC algorithm. Initialize with starting values by setting $s=0$, draw $%
d_{i}^{(0)}\sim \mathit{DiscreteUniform}(1,n)$ for $i=1,\cdots ,n$. Then for $%
j=1,\cdots ,M^{(0)}=\max\limits_{i}d_{i}^{(0)}$, draw $\lambda _{j}^{(0)}\sim \textit{Exp}(1)$, ${\gamma _{j}^{-2}}^{(0)}\sim \mathit{Ga}(2,\frac{2V}{%
	\lambda _{j}^{(0)}})$, where $V = \frac{1}{p} \sum\limits_{l=1}^p \hat{\beta}_{l}^2 \bm{1}(n \geq p+1) + \frac{1}{n} \sum\limits_{l=1}^p \tilde{\beta}_{l}^2 \bm{1}(n < p + 1)$. We also initialize starting values for $\psi _{jl}^{(0)}\sim \mathit{Ga}%
(\lambda _{j}^{(0)},2{\gamma _{j}^{-2}}^{(0)})$, $\mu _{j}^{(0)}\sim \N(0,\nu
_{\mu })$, $\bm{\beta}_{j}^{(0)}\sim \N_{p}(\bm{0}_{p},\bm{D}_{\psi
	,j}^{(0)}) $, $m_{jl}^{(0)}\sim \N(m_{0},\tau _{jl}^{(0)}/n_{0})$, $\tau
_{jl}^{(0)}\sim \mathit{IG}(\frac{\nu _{0}}{2},\frac{2}{\nu
	_{0}s_{0}^{2}})$, ${\sigma ^{2}}^{(0)} \sim \mathit{IG}(\alpha _{0},\theta_{0}) $, and mass parameter $\alpha ^{(0)}\sim \mathit{Ga}(\alpha _{\alpha
},\theta _{\alpha })$.

Then, for each iteration $s=1,\cdots ,S$, draw from the full conditional 
posterior distributions described in the following steps:

\textbf{Step 1: Draw mixture weights} $w_{j}^{(s)}$: For $j=1,\cdots ,M^{(s)}=\max\limits_{i}d_{i}^{(s-1)}$, take $%
n_j^{(s)}=\sum\limits_{i=1}^{n}\bm{1}(d_{i}^{(s-1)}=j)$, $%
m_{j}^{(s)}=\sum\limits_{i=1}^{n}\bm{1}(d_{i}^{(s-1)}>j)$. Then for $%
j=1,\cdots ,M^{(s)}$, draw $v_{j}^{(s)}\sim \mathit{Beta}(1+n_j^{(s)},\alpha
^{(s-1)}+m_{j}^{(s)})$, let $w_{j}^{(s)}=v_{j}^{(s)}\prod%
\limits_{l<j}(1-v_{j}^{(s)})$, and draw $u_{i}^{(s)}\sim \mathit{Unif}%
(0,w_{d_{i}^{(s-1)}}^{(s)})$ for $i=1,\cdots ,n$. For $j=M^{(s)}+1,\cdots
,N^{(s)}$, draw $v_{j}^{(s)}\sim \mathit{Beta}(1+n_j^{(s)},\alpha
^{(s-1)}+m_{j}^{(s)})$ until the smallest $N^{(s)}$ is obtained such that $%
\sum\limits_{j=1}^{N^{(s)}}w_{j}^{(s)}>\max\limits_{i}(1-u_{i}^{(s)})$.%

\textbf{Step 2}: For $j=1,\cdots ,M^{(s)}$, update $\lambda _{j}^{(s)}$, ${%
	\gamma _{j}^{-2}}^{(s)}$, $\bm{\psi}_{j}^{(s)}$, $\mu _{j}^{(s)},\bm{\beta}%
_{j}^{(s)}$, $\m_{j}^{(s)},\bm{\tau}_{j}^{(s)}$:

\textbf{2.1}: Draw from:
\begin{align*}
&\lambda _{j}^{(s)}|\bm{\psi}_{j}^{(s-1)},{\gamma _{j}^{-2}}%
^{(s-1)} \propto \pi (\lambda _{j})\prod\limits_{l=1}^{p}\pi (\psi
_{jl}^{(s-1)}|\lambda _{j},{\gamma _{j}^{-2}}^{(s-1)}) \propto \pi (\lambda _{j})(\frac{1}{2}{\gamma _{j}^{-2}}%
^{(s-1)})^{p\lambda _{j}}[\Gamma (\lambda _{j})]^{-p}\Big[%
\prod\limits_{l=1}^{p}\psi _{jl}^{(s-1)}\Big]^{\lambda _{j}}.
\end{align*}%
Since the conditional posterior distribution is not in closed form, we perform a sampling
update of $\lambda _{j}$ using the stepping-out slice sampling algorithm %
\citep{neal2003slice}.

\textbf{2.2}: Draw from:
\begin{align*}
&& {\gamma _{j}^{-2}}^{(s)}|\lambda _{j}^{(s)},\bm{\psi}_{j}^{(s-1)} &\propto
\pi (\gamma _{j}^{-2}|\lambda _{j}^{(s)})\prod\limits_{l=1}^{p}\pi (\psi
_{jl}^{(s-1)}|\lambda _{j}^{(s)},\gamma _{j}^{-2}) \propto \mathit{Ga}(\gamma _{j}^{-2}|2,V/2\lambda _{j}^{(s)} )\prod\limits_{l=1}^{p}\Big[(2\gamma_{j}^{-2})^{\lambda }\text{exp}\{-\frac{1}{2}\gamma_{j}^{-2}\psi_{jl}^{(s-1)}\}\Big] \\
&&& \sim \mathit{Ga}\Big(p\lambda _{j}^{(s)}+2,\Big[\frac{1}{2}%
\sum\limits_{l=1}^{p}\psi _{jl}^{(s-1)}+V/2\lambda _{j}^{(s)}\Big]\Big).
\end{align*}

\textbf{2.3}: For $l=1,\cdots ,p$, draw from the Generalized Inverse Gaussian (\textit{GIG}) distribution, where $\textit{GIG}(c,d,h)$ has the probability density\\ $f(x) = \frac{(c/d)^{h/2}}{2K_h(\sqrt{cd})} x^{h-1} e^{(cx+d/x)/2}$.
\begin{align*}
&& \psi _{jl}^{(s)}|\beta _{jl}^{(s-1)},\lambda _{j}^{(s)},{\gamma _{j}^{-2}%
}^{(s)} & \propto \pi (\psi _{jl}|\lambda _{j}^{(s)},{\gamma _{j}^{-2}}%
^{(s)})\pi (\beta _{jl}^{(s-1)}|\psi _{jl}) \propto \psi _{jl}^{\lambda _{j}^{(s)}-1}\text{exp}\{-\frac{1}{2}{\gamma
	_{j}^{-2}}^{(s)}\psi _{jl}\}\psi _{jl}^{-\frac{1}{2}}\text{exp}\{-\frac{1}{%
	2\psi _{jl}}{\beta _{jl}^{2}}^{(s-1)}\} \\
&&& \sim \textit{GIG}\Big(\lambda _{j}^{(s)}-\frac{1}{2},{\gamma _{j}^{-2}}%
^{(s)},{\beta _{jl}^{2}}^{(s-1)}\Big).
\end{align*}

\textbf{2.4}: If $n_j>p+1$, draw from:
\begin{align*}
\mu _{j}^{(s)},\bm{\beta}_{j}^{(s)}|\bm{D}_{\psi ,j}^{(s)},{\sigma ^{2}}%
^{(s-1)} &\propto \prod\limits_{d_{i}^{(s-1)}=j}\N(y_i|\x_i,\mu _{j},%
\bm{\beta}_{j},{\sigma ^{2}}^{(s-1)})\pi (\bm{\beta}_{j}|\bm{D}_{\psi
	,j}^{(s)})\pi (\mu _{j}) \\
& \sim \N_{p+1}(B^{-1}{\tilde{\X}_j}^{*^T} \bm{\y}_j^*,{\sigma ^{2}}%
^{(s-1)}B^{-1}),
\end{align*}%
where $B={\tilde{\X}_j}^{*^T} {\tilde{\X}_j}^{*} +{\sigma ^{2}}%
^{(s-1)}\Lambda _{j}^{(s)}$, ${\tilde{\X}_j}^{*^T} =[\bm{1}_{n_j},\X_j^*]$, $\Lambda
_{j}^{(s)}=\text{diag}\Big(0,1/\psi _{j1}^{(s)},\cdots ,1/\psi _{jp}^{(s)}%
\Big)$. Otherwise, if $n_{j}\leq p+1$, we take the singular value decomposition 
${\tilde{\X}_j}^{*} =F^{T}DA^{T}$, and let $\hat{\theta}%
_{n_{j}}=D^{-1}F \bm{\y}_j^*$. Then we draw from:
\begin{align*}
\mu _{j}^{(s)},\bm{\beta}_{j}^{(s)}|\bm{D}_{\psi ,j}^{(s)},{\sigma ^{2}}%
^{(s-1)}\sim \N_{p+1}(\Psi _{j}^{(s)}AC^{-1}\hat{\theta}_{n_{j}},\Psi
_{j}^{(s)}-\Psi _{j}^{(s)}AC^{-1}A^{T}\Psi _{j}^{(s)})
\end{align*}%
where $\Psi _{j}^{(s)}=\text{diag}\Big(\nu _{\mu },\psi _{j1}^{(s)},\cdots
,\psi _{jp}^{(s)}\Big)$, $C=\Psi _{0_{j}}+{\sigma ^{2}}^{(s-1)}\Lambda ^{\ast
}$, $\Psi _{0_{j}}=A^{T}\Psi _{j}^{(s)}A$, $\Lambda ^{\ast }=D^{-2}$.%

\textbf{2.5}: For $l=1,\dots ,p$, draw from:
\begin{align*}
\tau _{jl}^{(s)}\propto \prod\limits_{d_{i}^{(s-1)}=j} \N(x_{il}|m_{jl}^{(s-1)},\tau _{jl})\pi (\tau _{jl})\sim \mathit{IG}(\frac{%
	\nu ^{\ast }}{2},\frac{2}{{s^{\ast }}^{2}\nu ^{\ast }}),
\end{align*}%
where $n^{\ast }=n_{0}+n_{j}$, $\nu ^{\ast }=\nu _{0}+n_{j}$, $\bar{x}_{n_{j},l}=\frac{1}{n_{j}}\sum\limits_{i=1}^{n_{j}}x_{il}$, and ${%
	s^{\ast }}^{2}=\frac{1}{\nu ^{\ast }}\Big[\sum\limits_{i=1}^{n_{j}}(x_{il}-\bar{x}_{n_{j},l})^{2}+s_{0}^{2}\nu _{0}+\frac{n_{0}n_{j}}{%
	n^{\ast }}(\bar{x}_{n_{j},l}-m_{0})^{2}\Big]$.

\textbf{2.6}: For $l=1,\dots ,p$, draw from:
\begin{align*}
m_{jl}^{(s)}|\tau _{jl}^{(s)}\propto \prod\limits_{d_{i}^{(s-1)}=j} \N(x_{il}|m_{jl},\tau _{jl}^{(s)})\pi (m_{jl}|\tau _{jl}^{(s)})\sim \N(m^{\ast
},\frac{\tau _{jl}^{(s)}}{n^{\ast }}),
\end{align*}%
where $m^{\ast }=(n_{j}\bar{x}_{n_{j},l}+n_{0}m_{0}) / (n_{j}+n_{0})$.

If $N^{(s)}>M^{(s)}$, then for $j=M^{(s)}+1,\cdots ,N^{(s)}$ , draw ${	\lambda _{j}}^{(s)}\sim \textit{Exp}(1)$, ${\gamma _{j}^{-2}}^{(s)}\sim \textit{Ga}(2,V/2\lambda _{j}^{(s)})$, $\bm{\psi}_{j}^{(s)}\sim
\prod\limits_{l=1}^{p}\mathit{Ga}(\lambda _{j}^{(s)},{\gamma _{j}^{-2}}%
^{(s)})$, $\bm{\beta}_{j}^{(s)}\sim \N(\bm{0}_{p},\bm{D}_{\psi ,j}^{(s)})$, $%
\mu _{j}^{(s)}\sim \N(0,\nu _{\mu })$, $\bm{\tau}_{j}^{(s)}\sim
\prod\limits_{l=1}^{p}\mathit{IG}(\nu _{0}/2,2/\nu _{0}s_{0}^{2})$, $%
\m_{j}^{(s)}\sim \prod\limits_{l=1}^{p}\N(m_{0},\tau _{jl}^{(s)}/n_{0})$. 

\textbf{Step 3}: For $i=1,\cdots,n$, sample $%
d_i^{(s)} = j$ with probability proportional to:
\begin{align*}
\bm{1}(w_j^{(s)}>u_i^{(s)}) \N(y_i | \mu_j^{(s)}+\x_i^T\bm{\beta}_j^{(s)},{%
	\sigma^2}^{(s-1 )}) \N_p(\x_i|\m_j^{(s)}, \bm{\tau}_j^{(s)}) , \text{ for } j=1,\cdots,N^{(s)}.
\end{align*}

\textbf{Step 4}: Draw from:
\begin{align*}
{\sigma^2}^{(s)} | \bm{\beta}^*_k, \mu^*_k &\propto \prod\limits_{i=1}^n \N(y_i | \mu_{d_i}^{(s)}+\x_i^T\bm{\beta}_{d_i}^{(s)},{\sigma^2}) \pi(\sigma^2)
\\
& \sim \mathit{IG}\Big( n/2 + \alpha_0, \sum\limits_{i=1}^n(y_i -
\mu_{d_i}^{(s)} - \x_i^T \bm{\beta}_{d_i}^{(s)})^2 / 2 + \theta_0 \Big).
\end{align*}

\textbf{Step 5}: Update $\alpha^{(s)}$ by drawing from a Gamma distribution
with shape $\alpha_{\alpha} + K - \bm{1}(u > \{O/(1+O)\})$ and scale $%
\theta_{\alpha}-\log(\eta)$, where $K$ is number of unique $d_i^{(s)}$'s, $\eta
\sim \mathit{Beta}(\alpha^{(s-1)}+1, n)$, $u \sim \textit{Unif}(0,1)$ and $O
= (\alpha_{\alpha} + K - 1)/( \{\theta_{\alpha} - \log(\eta)\} n)$ \citep{escobar1995bayesian}.

Sampling updates from Steps 1 through 5 are repeated for a large number of iterations until the MCMC chain has displayed good mixing, according to trace plots.

The Python code and the R code files, provide more details about the MCMC sampling algorithm, along with the MCMC code files for the N-DPM model, and the simulated data sets and real data set analyzed in Sections 3 and 4 of the article.

\end{document}